# Extremely Thin Dielectric Metasurface for Carpet Cloaking

LiYi Hsu, Thomas Lepetit, and Boubacar Kanté[*]

**Abstract**-We demonstrate a novel and simple approach to cloaking a scatterer on a ground plane. We use an extremely thin dielectric metasurface ($\lambda/12$) to reshape the wavefronts distorted by a scatterer in order to mimic the reflection pattern of a flat ground plane. To achieve such carpet cloaking, the reflection angle has to be equal to the incident angle everywhere on the scatterer. We use a graded metasurface and calculate the required phase gradient to achieve cloaking. Our metasurface locally provides additional phase to the wavefronts to compensate for the phase difference amongst light paths induced by the geometrical distortion. We design our metasurface in the microwave range using highly sub-wavelength dielectric resonators. We verify our design by full-wave time-domain simulations using micro-structured resonators and show that results match theory very well. This approach can be applied to hide any scatterer on a ground plane not only at microwave frequencies, but also at higher frequencies up to the near infrared.

## 1. Introduction

Due to their amazing ability to manipulate electromagnetic waves, metamaterials have been extensively studied in the past fifteen years. They have resulted in several novel concepts and promising applications, such as cloaking devices [1-12], concentrators [13-15], wormholes [16] and hyperlenses [17]. Among all potential applications, invisibility cloaks especially received considerable attention.

Up to now, the main theoretical tool used for designing invisibility cloaks has been transformation optics [1-2]. According to Fermat's principle, an electromagnetic wave will travel between two points along the path of least time. In a homogeneous material, this path is just a straight line. However, in an inhomogeneous material, the path becomes a curve because waves travel at different speeds at different points. Thus, one can control the path of waves by appropriately designing the material parameters (electric permittivity and magnetic permeability). In the case of cloaking, a metamaterial surrounding the target can be used to force light bypass a region of space, effectively isolating it from incoming electromagnetic waves.

Using transformation optics as the design method, the first experimental demonstration of cloaking was achieved at microwave frequencies [3]. However, transformation optics usually leads to highly anisotropic and inhomogeneous material parameters. Besides, extreme material parameter values, such as negative or near-zero, are also often required. To obtain extreme values for the permeability, split-ring resonators (SRRs) with magnetic resonances have been used. Such resonances are strongly dispersive and result in a cloak working only in a narrow frequency range. Most metals are also highly lossy at optical frequencies, which prohibits a simple scaling of SRRs down to the nanoscale [18-21].

[*] Corresponding author: B. Kanté (bkante@ucsd.edu)
The authors are with the Department of Electrical and Computer Engineering, University of California San Diego, La Jolla, California 92093-0407, USA.

Recently, a refinement of the transformation optics strategy was put forward [4]. It is called 'hiding under the carpet' and works not by routing light around a given scatterer but by transforming its reflection pattern into that of a flat plane. With a well-designed metamaterial, reflected waves appear to be coming from a flat plane and the scatterer on top of it thus becomes invisible. Still, a major drawback of metamaterial-based cloaking devices is that they are large in size. This is because a large space is needed to progressively bend light.

Recently, metasurfaces have been used to overcome this specific disadvantage of metamaterials, although in a different context [22-23]. Metasurfaces or frequency selective surfaces, which are the surface version of metamaterials, have the inherent advantage of taking up less physical space than metamaterials. They have raised significant attention due to the simplified design afforded by generalized Snell's laws of reflection and refraction [24]. In reference 24, it was shown in a simple and elegant way that wave propagation can be controlled by using a thin coating layer with a properly designed phase gradient over the surface.

We propose in this paper a dielectric metasurface with a tailored phase gradient for carpet cloaking. We show that a single extremely thin all-dielectric metasurface is sufficient to accomplish invisibility. Once the scatterer is covered with the designed metasurface, no observer can distinguish it from a flat surface.

## 2. Phase Distribution

The carpet surface under which cloaking is achieved is described by a curve y(x,z) in 2D. To illustrate our design strategy, we consider a scattering surface (invariant in z) that is described by a 1D Gaussian function.

$$y(x) = Ae^{-\frac{x^2}{\sigma^2}} \quad (1)$$

To illustrate the cloaking mechanism, we consider two simple cases. In Fig. 1a, an incident wave is reflected by a flat ground plane. Snell's law dictates that the reflection angle is equal to the incident angle ($\theta_r = \theta_i$). In Fig. 1b, when the flat ground plane is rotated counterclockwise by an angle $\varphi$, the new incident angle becomes $\theta_i - \varphi$ while the new reflection angle becomes $\theta_r + \varphi$. Approximating each point of the Gaussian scatterer locally by a flat plane, we can design the entire cloak simply based on the geometric considerations made in Figs. 1a-b.

To control the reflection angle, we use the generalized Snell's law of reflection [24]

$$sin(\theta_r) - sin(\theta_i) = \frac{1}{k_i}\frac{d\Phi(x)}{dx} \quad (2)$$

We have introduced the wavevector in the incident medium $k_i$ and the phase distribution $\Phi(x)$. From Eq. (2), the reflection angle is entirely controlled by the phase gradient. Various phase gradients can be achieved with a graded metasurface. For example, we can design a suitable phase gradient on the plane to ensure that the reflected ray in Fig. 1b follows the same path as the one in Fig. 1a. Hence, the observer will be lead to believe that he/she sees the original flat ground plane without any rotation.

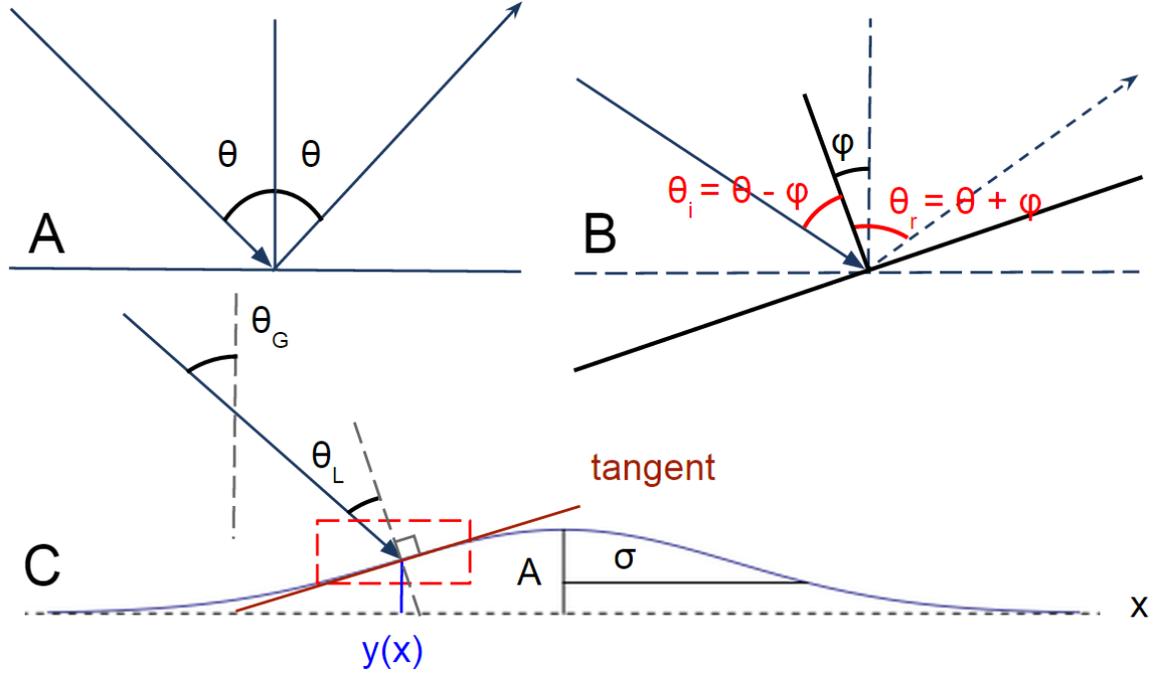

Fig. 1: (A) Reflection from a flat plane. (B) Reflection from a flat plane with a counterclockwise rotation by an angle φ. Cases A and B are both governed by Snell's law. (C) Reflection from a Gaussian scatterer that can be treated locally, at each point along the surface, as a flat plane.

Treating each point on the Gaussian scatterer locally as a flat plane, we can parametrize the entire scatterer by a local incident angle $\theta_L$ that is x-dependent and that is distinct from the global incident angle $\theta_G$ (see Fig. 1c). Assuming the wave is propagating in vacuum we can then write

$$sin(2\theta_G - \theta_L) - sin(\theta_L) = \frac{1}{k_0}\frac{d\Phi(x)}{dx} \quad (3)$$

The phase gradient can then be expressed as a function of the scatterer's shape $y(x)$

$$\frac{d\Phi(x)}{dx} = 2k_0 cos\theta_G \frac{\frac{dy(x)}{dx}}{\sqrt{1 + \left(\frac{dy(x)}{dx}\right)^2}} \quad (4)$$

Finally, after integration the phase distribution $\Phi(x)$ is given by

$$\Phi(x) = 2k_0 cos\theta_G \int \frac{\frac{dy(x)}{dx}}{\sqrt{1 + \left(\frac{dy(x)}{dx}\right)^2}} dx \quad (5)$$

From Eq. (5), we see immediately that in the limit of a flat scatterer, the phase distribution is identically constant as it should be. By providing the appropriate phase distribution, as dictated by Eq. 5, we can hide an arbitrary scatterer by making it look like a flat ground plane. Clearly, the concept introduced here can be realized at any frequency by simply picking a proper class of sub-wavelength metasurface elements.

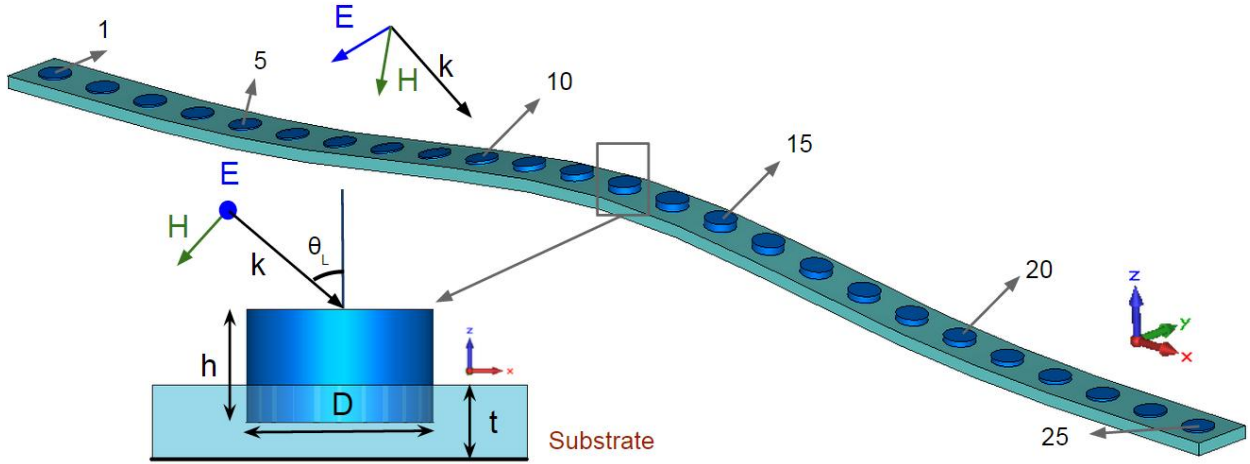

Fig. 2: Schematic of the entire metasurface, discretized with 25 cylinders, and coordinate system. (Inset) Unit cell of the metasurface. The bottom black line is the ground plane, the light blue substrate is Teflon, and the dark blue cylinder is ceramic. The incident wave is TM-polarized.

## 3. Dielectric Metasurface

We design a microwave metasurface made of dielectric cylinders for a frequency of 4.15 GHz (C-band). Cylinders have a circular cross-section and a fixed diameter ($D = 0.58$ in) and the substrate has a fixed thickness ($t = 0.23$ in). Our metasurface is periodic along $y$ with a sub-wavelength unit cell ($w = 1.16$ in). Cylinders are made of a high-permittivity ceramic ($\varepsilon_r = 41 \pm 0.75$) with a low loss tangent ($tan\delta = 1.10^{-4}$) and are embedded in a Teflon substrate ($\varepsilon_r = 2.1$) with an equally low loss tangent ($tan\delta = 2.10^{-4}$). Our metasurface is thus almost lossless.

The scatterer is described by a Gaussian function as per Eq. 1. Its standard deviation $\sigma$ is four times the unit cell width ($\sigma = 4.64$ in) while its amplitude $A$ is the same as the unit cell width ($A = 1.16$ in). Finally, the global incident angle $\theta_G$ is chosen to be 45 degrees and the polarization of the incident wave is parallel with the surface of the cylinder (*i.e.*, TM-polarized).

To obtain a suitable phase gradient and phase distribution, we design a local variation in cylinder height. Note that this is the only geometrical parameter that is varied. As shown in Fig. 3, from the scatterer geometry $y(x)$, we compute the local incident angle $\theta_L(x)$, and then the phase distribution $\Phi(x)$ from Eq. 5. As can be seen from Table 1, to hide the Gaussian scatterer phase covering the 0-to-$2\pi$ range is needed for different local incident angles.

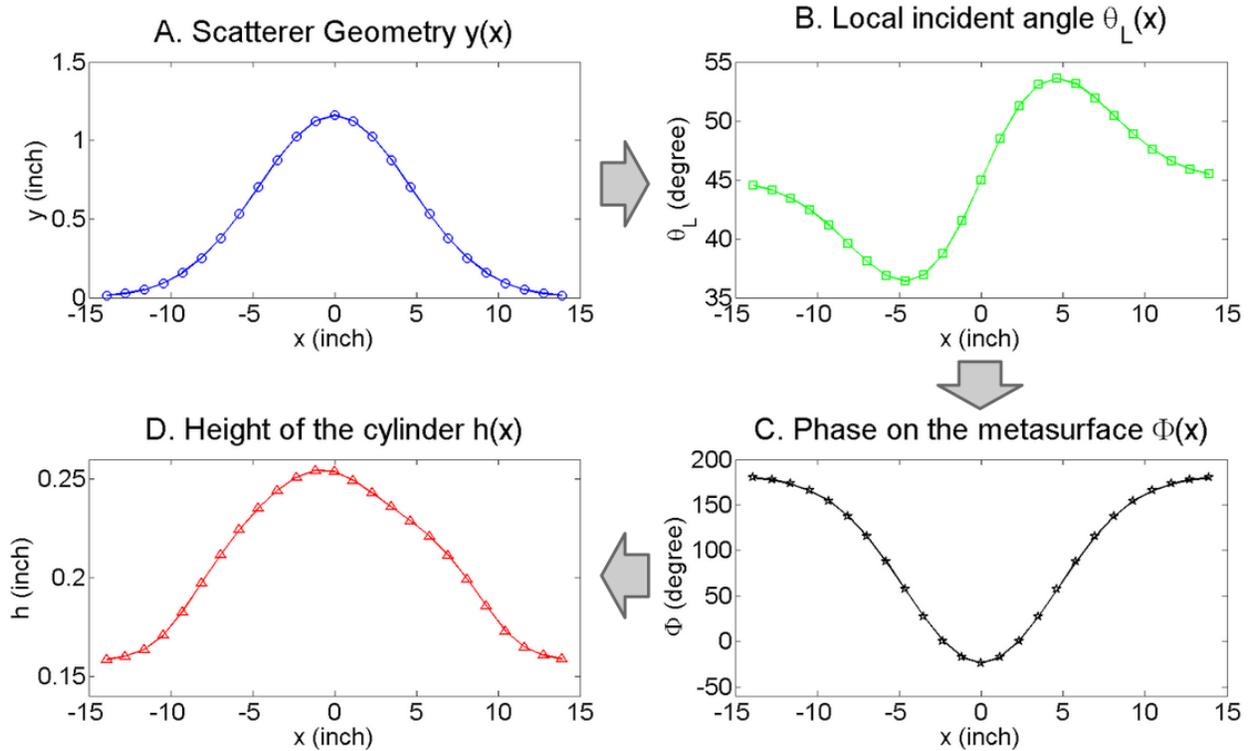

Fig. 3: Graded metasurface design flowchart (A) Scatterer geometry vs. position x. (B) Local incident angle vs. position x. (C) Phase shift vs. position x. (D) Height of the cylinder vs. position x.

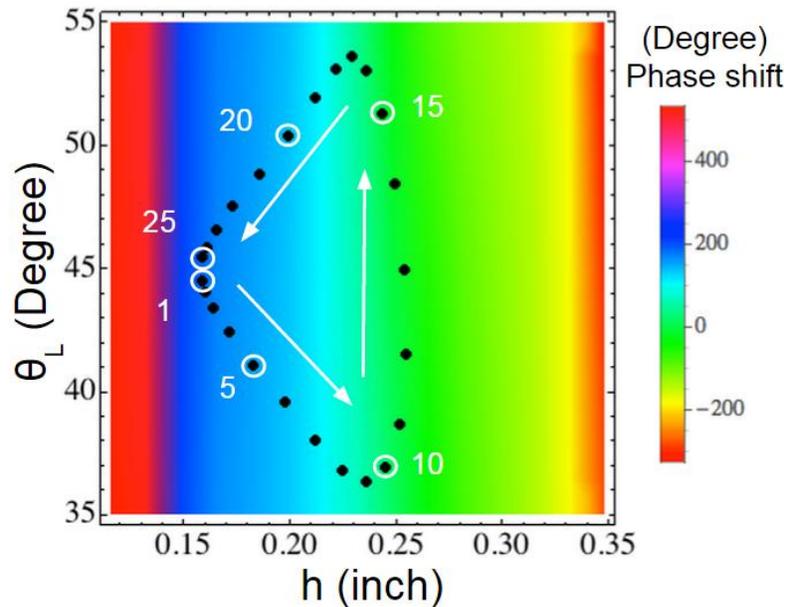

Fig. 4: Simulated phase shift with varying height $h$ and local incident angle $\theta_L$ for a frequency of 4.15 GHz. The dark points correspond to the different heights chosen for the 25 cylinders on the metasurface.

| Function \ Index | 1 | 5 | 10 | 15 | 20 | 25 |
|---|---|---|---|---|---|---|
| y (in) | 0.01 | 0.16 | 0.88 | 1.02 | 0.25 | 0.01 |
| $\theta_L$ (deg) | 44.5 | 41.1 | 36.9 | 51.3 | 50.4 | 45.5 |
| Φ (deg) | 180.0 | 154.2 | 26.7 | 0.4 | 137.5 | 180.0 |
| h (in) | 0.16 | 0.18 | 0.24 | 0.24 | 0.20 | 0.16 |

Table. 1: Samples of calculated y(x), $\theta_L$(x), Φ(x) and h(x) on the scatterer.

To see if the required phase coverage is achievable for different local incident angles $\theta_L$ with our dielectric cylinders, we simulate the phase shift as a function of both local incident angle and cylinder height. As can be seen from Fig. 4, the phase varies over more than $2\pi$ for the entire range of local incident angles required ($35° \leq \theta \leq 55°$). By interpolating the $\theta_L - h$ diagram in Fig. 4, we obtain the height needed for each dielectric cylinder, i.e., $h(x)$. We discretize the phase distribution with 25 cylinders.

To compute the phase shift from a single metasurface element we assume that its response can be approximated by that of an infinitely periodic array. In our case, this is a good approximation because cylinders are made of a high permittivity material that concentrates the field and, as a result, the coupling between unit cells is weak enough to consider each unit cell as independent. Furthermore, since the phase gradients are small, nearby cylinders are of comparable dimensions. Thus, the total field of the whole system can be treated as the superposition of each unit cell as follows from Huygens principle.

## 4. Full Cloak Simulation

Based on the design from the previous section, we model the structure in a commercial full-wave solver [25], as shown in Fig. 2. Figures 5a-c show the electric field reflection pattern, i.e., the pattern obtained by subtracting the incident electric field from the total electric field, for the ground plane, the Gaussian scatterer, and the Gaussian scatterer covered by the cloaking metasurface, respectively. Figure 5d is a phase plot along the equiphase line L in Fig. 5a-c.

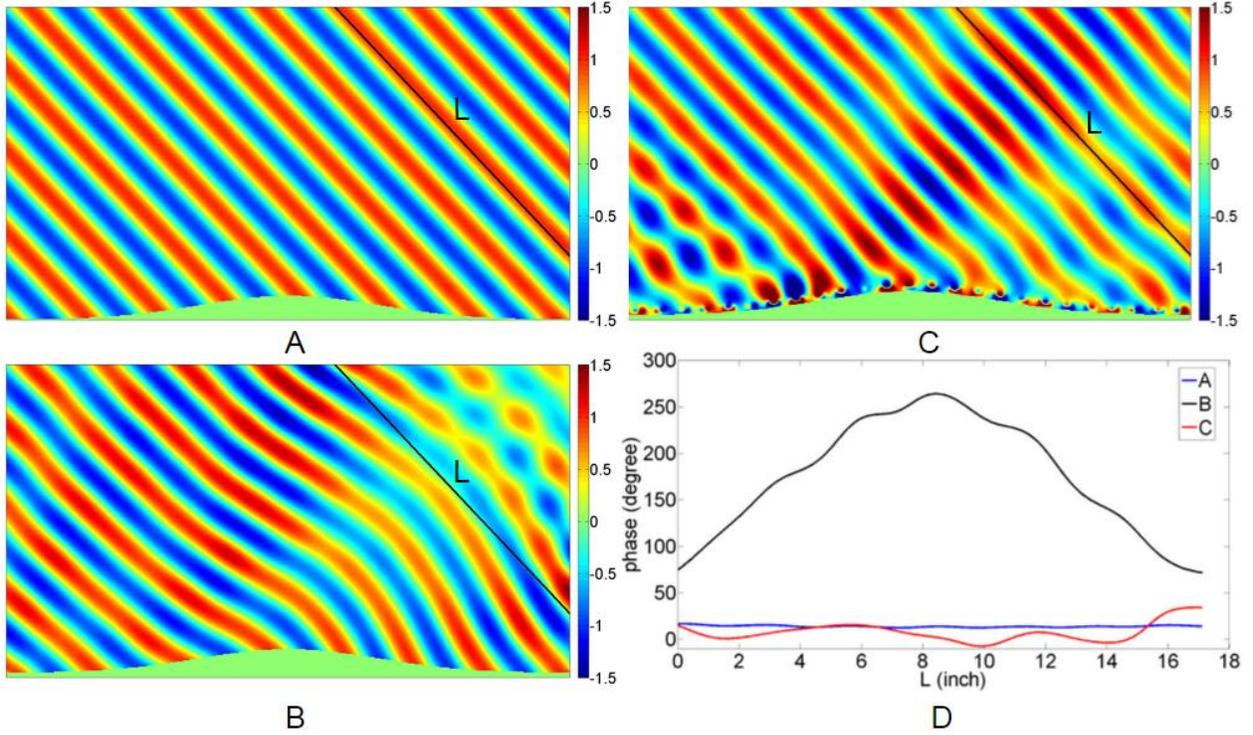

Fig. 5 (A) Electric field reflection pattern for a flat ground plane. (B) Electric field reflection pattern for a Gaussian scatterer. (C) Electric field reflection pattern for a Gaussian scatterer with cloaking metasurface. (D) Phase plot along the equiphase line L in (A)-(C).

In Fig. 5b, we observe the expected distortion due to the scatterer and in Fig. 5c its correction as provided by the metasurface. It is clear that the metasurface fixes the distortion considerably and the reflection pattern is that of a quasi-plane wave. Even with just about two cylinders per wavelength, we achieve a very good level of reflection pattern.

A sensitivity analysis can be carried out by computing the partial derivatives with respect to $x$, $\theta$, and $k_0$.

$$d\Phi(x, \theta, k_0) = \frac{\partial \Phi}{\partial x} dx + \frac{\partial \Phi}{\partial \theta} d\theta + \frac{\partial \Phi}{\partial k_0} dk_0 \qquad (6)$$

From Eqs. 5-6, we can draw three conclusions. First, the phase distribution sensitivity with respect to frequency is independent of frequency itself. There are thus no special considerations for different frequency ranges. Second, the phase distribution sensitivity with respect to global incident angle is maximum for grazing incidence ($\theta = \pi/2$). It is thus harder to cloak a scatterer for large angles of incidence. Finally, the phase distribution sensitivity with respect to position is, somewhat surprisingly, independent of position itself for large slopes.

All of this implies that a cloaking device can work for a large range of global incident angles and be broadband if the phase distribution on the metasurface is linear with respect to frequency and cosine-like with respect to global incident angle.

## Conclusion

In this paper, we have shown that carpet cloaking with an extremely thin dielectric metasurface is possible. Furthermore, the scheme we presented is general and can be extended to any shape and frequency range. By following our proposed design flow chart, one can easily design a metasurface cloak for a given geometry.

A specific design has been presented and cloaking performance has been shown to be robust with respect to surface discretization. The observed wavefronts reflected from the proposed metasurface have been shown to be quasi-planar, with little to no distortion. With this design, any observer will just see a flat ground plane and the scatterer will be invisible and thus effectively cloaked.

Moreover, the approach of bending electromagnetic waves can be used not only for carpet cloaks but also for light focusing to make flat optics devices such as thin solar concentrators quarter-wave plates and spatial light modulators.